\documentclass{article}

\PassOptionsToPackage{numbers, compress}{natbib}

\usepackage[final]{neurips_2023}

\usepackage[utf8]{inputenc} 
\usepackage[T1]{fontenc}    
\usepackage{hyperref}       
\usepackage{url}            
\usepackage{booktabs}       
\usepackage{amsfonts}       
\usepackage{nicefrac}       
\usepackage{microtype}      
\usepackage{xcolor}         

\usepackage{graphicx}
\usepackage{pifont}
\newcommand{\cmark}{\ding{51}}
\newcommand{\xmark}{\ding{55}}

\title{Pixels Under Pressure: Exploring Fine-Tuning Paradigms for Foundation Models in High-Resolution Medical Imaging}

\author{%
  Zahra ~TehraniNasab\\
  McGill University\\
  MILA-Quebec AI Institute\\
  \texttt{zahra.tehraninasab@mail.mcgill.ca} \\
  \And
  Amar ~Kumar\\
  McGill University\\
  MILA-Quebec AI Institute\\
  \texttt{amar.kumar@mail.mcgill.ca} \\
  \And
  Tal ~Arbel\\
  McGill University\\
  MILA-Quebec AI Institute\\
  \texttt{tal.arbel@mcgill.ca} \\
}

\begin{document}
\maketitle
\begin{abstract}
Advancements in diffusion-based foundation models have improved text-to-image generation, yet most efforts have been limited to low-resolution settings. As high-resolution image synthesis becomes increasingly essential for various applications, particularly in medical imaging domains, fine-tuning emerges as a crucial mechanism for adapting these powerful pre-trained models to task-specific requirements and data distributions. In this work, we present a systematic study, examining the impact of various fine-tuning techniques on image generation quality when scaling to high resolutions ($512\times512$ pixels). We benchmark a diverse set of fine-tuning methods, including full fine-tuning strategies and parameter-efficient fine-tuning (PEFT). We dissect how different fine-tuning methods 
influence key quality metrics, including Fr\'echet Inception Distance (FID), Vendi score, and prompt-image alignment. 
We also evaluate the utility of generated images in a downstream classification task under data-scarce conditions, demonstrating that specific fine-tuning strategies improve both generation fidelity and downstream performance when synthetic images are used for classifier training and evaluation on real images. Our code is accessible through the project website \footnote{\href{https://tehraninasab.github.io/PixelUPressure/}{https://tehraninasab.github.io/PixelUPressure/}}.
\end{abstract}

\section{Introduction}

Text-to-Image Foundation models have demonstrated remarkable success across various computer vision tasks, consistently achieving strong performance on standard benchmarks~\cite{li2024cosmicman,dong2024internlm}. Trained on massive corpora of natural images, these models acquire visual representations, such as textures, shapes, and complex spatial patterns, that often transfer effectively to medical imaging domains. This transferability is particularly valuable in clinical settings, where the availability of labeled or paired text-image medical data is scarce. Given the challenges of training large-scale models from scratch with small, specialized datasets, fine-tuning has emerged as a practical and effective strategy for adapting foundation models to medical applications~\cite{yu2024curriculum,liu2024pefomed,ruffini2025benchmarking}. Fine-tuning leverages the rich visual priors learned during pretraining, enabling models to be adapted to domain-specific tasks through targeted updates, rather than full retraining~\cite{zhang2024challenges,azad2023foundational}. In this work, we focus on Stable Diffusion~\cite{Rombach_2022_CVPR} v1.5, a prominent latent diffusion model pre-trained on large-scale natural image-text pairs. While its latent-space architecture enables more efficient high-resolution synthesis compared to pixel-space alternatives, fine-tuning Stable Diffusion on high-resolution medical images still poses significant computational challenges. Scaling to high resolution rapidly increases memory and compute costs, especially for attention-based architectures, posing significant barriers to deployment~\cite{zhang2024challenges}. These challenges introduce a trade-off between model capacity and computational feasibility, often leading to compromises in generative quality or diagnostic reliability. Addressing this tension is essential for enabling the practical and scalable use of text-to-image foundation models in real-world medical imaging workflows~\cite{dutt2023parameter}.

Recent developments in parameter-efficient fine-tuning (PEFT) methods such as Low-Rank Adaptation (LoRA)~\cite{hu2022lora}, DoRA (Decoupled Rank Adaptation)~\cite{liu2024dora}, BitFit~\cite{zaken2021bitfit} and Diffusion-specific PEFT (DiffFit)~\cite{xie2023difffit}, offer promising solutions to these computational scaling challenges by selectively updating model components rather than the entire parameter space~\cite{dutt2023parameter}.  Adapter modules~\cite{li2022cross,chen2022adaptformer} offer an alternative approach by inserting small, trainable layers between frozen, pre-trained components, enabling domain-specific adaptation without modifying the original model weights.  
Despite these methodological advances and their demonstrated effectiveness on standard resolution tasks, the performance characteristics and scaling behaviour of these parameter-efficient approaches in high-resolution medical imaging contexts ($512\times512$) remain largely unexplored, particularly regarding their ability to preserve critical diagnostic information while maintaining computational efficiency as input dimensions increase substantially. Previous works by Dutt et al.~\cite {dutt2023parameter} have analyzed the effect of different strategies to fine-tune at $224\times224$ resolution but did not focus on the effect of fine-tuning on image generation quality at higher resolution. Recent work by Davila et al.~\cite{davila2024comparison} have compared fine-tuning strategies for image classification in medical imaging, but these images are also at a low resolution of $320\times320$ pixels.

This paper presents a comprehensive study of fine-tuning paradigms for high-resolution image generation using Stable Diffusion, a pre-trained diffusion-based text-to-image foundation model. The key contributions of our work are as follows:
\begin{itemize}
    \item A systematic comparison of fine-tuning strategies---including full fine-tuning and parameter-efficient approaches such as LoRA~\cite{hu2022lora}, DoRA~\cite{liu2024dora}, BitFit~\cite{zaken2021bitfit}, and DiffFit~\cite{xie2023difffit}---for $512\times512$ image synthesis.

    \item An in-depth analysis of how these fine-tuning strategies affect generation quality, including visual fidelity (measured by Fr'echet Inception Distance, FID~\cite{heusel2017gans}), diversity (via Vendi Score~\cite{friedman2022vendi}), and prompt-image consistency (evaluated using a classifier-based metric).

    \item A downstream evaluation where classifiers are trained on synthetic images and tested on real data, assessing the utility of generated images for real-world diagnostic tasks.
\end{itemize}

\section{Methodology}
\begin{figure}[t]
    \centering
    \includegraphics[width=\textwidth]{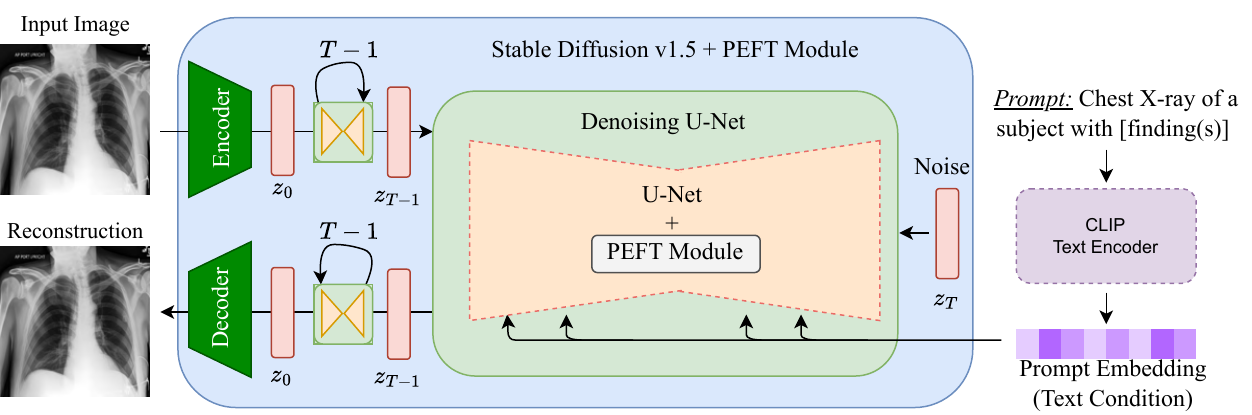}
    \caption{
    Overview of our architecture. Stable Diffusion v1.5 model is adapted for high-resolution chest X-ray generation using different fine-tuning strategies. 
    The PEFT module shown within the U-Net is only used in parameter-efficient fine-tuning configurations (e.g., LoRA, DoRA, BitFit); in full fine-tuning settings, the U-Net (and optionally the VAE and text encoder) are directly fine-tuned without PEFT modules.
    }
    \label{fig:architecture}
\end{figure}

\subsection{Fine-tuning Strategies}
We evaluate diverse fine-tuning configurations (Figure~\ref{fig:architecture}) for high-resolution image generation through an extensive set of experiments covering combinations of VAE, U-Net, and text encoder using full and parameter-efficient fine-tuning (PEFT); see Table~\ref{tab:fine-tuning-strategies}. Following~\cite{kumar2025prism,tehraninasab2025language}, we generate image-text pairs from tabular data using the template  -  \texttt{Chest X-ray of a subject with [disease(s)\footnote{The diseases include: \texttt{No Finding}, \texttt{Enlarged Cardiomediastinum}, \texttt{Cardiomegaly}, \texttt{Lung Opacity}, \texttt{Lung Lesion}, \texttt{Edema}, \texttt{Consolidation}, \texttt{Pneumonia}, \texttt{Atelectasis}, \texttt{Pneumothorax}, \texttt{Pleural Effusion}, \texttt{Pleural Other}, \texttt{Fracture}, and \texttt{Support Devices}.}]}. 
\begin{table}[ht]
\centering
\caption{Summary of fine-tuning strategies. \xmark: Frozen, \cmark: Trainable}
\begin{tabular}{@{} c c c c c @{}}
\toprule
\textbf{\#Model}
  & \textbf{VAE}
  & \textbf{Text Encoder}
  & \textbf{U-Net}
  & \textbf{Description / Trainable} \\
\midrule
\multicolumn{5}{l}{\itshape A. Full Fine-Tuning Strategies}\\
1  & \xmark & \xmark & \cmark 
   & U-Net only  \\
2  & \xmark & \cmark & \xmark 
   &  Text Encoder only   \\
3  & \cmark & \xmark & \xmark 
   & VAE only   \\
4  & \xmark & \cmark & \cmark 
   & Text encoder + U-Net  \\
5  & \cmark & \xmark & \cmark 
   & VAE + U-Net  \\
6  & \cmark & \cmark & \xmark 
   &  VAE + Text Encoder  \\
7  & \cmark & \cmark & \cmark 
   & U-Net + VAE + Text Encoder \\
\addlinespace
\multicolumn{5}{l}{\itshape B. Parameter-Efficient Fine-Tuning on U-Net}\\
8  & \xmark & \xmark & \cmark 
   & LoRA \cite{hu2022lora} \\
9  & \xmark & \xmark & \cmark 
   & DoRA \cite{liu2024dora} \\
10 & \xmark & \xmark & \cmark 
   & BitFit: Only bias terms updated \cite{zaken2021bitfit} \\
11 & \xmark & \xmark & \cmark 
   & DiffFit: Diffusion-specific method \cite{xie2023difffit} \\
\bottomrule
\end{tabular}
\label{tab:fine-tuning-strategies}
\end{table}

\noindent \textbf{Full Component Fine-Tuning}
For the full component fine-tuning experiments (Models 1–7), we explored the effects of selectively training different combinations of the VAE, Text Encoder, and U-Net modules. This approach allowed us to isolate the contribution of each component to the overall performance of the diffusion model. Some models focused on fine-tuning a single component while freezing the others to assess its standalone impact. For example, Model 1 fine-tuned only the U-Net, while Models 2 and 3 focused solely on the Text Encoder and VAE, respectively. Others involved combinations of two components (Models 4–6) or all three (Model 7).

\noindent \textbf{Parameter Efficient Fine-Tuning}
To improve efficiency and deployment flexibility, we explored four prominent parameter-efficient fine-tuning strategies:
\begin{itemize}
    \item \textit{Low-Rank Adaptation (LoRA)}~\cite{hu2022lora}: Uses Low-Rank Adaptation to insert trainable low-rank matrices into the network layers, significantly reducing trainable parameters while retaining adaptability.
    
    \item \textit{Decoupled Rank Adaptation (DoRA)}~\cite{liu2024dora}: Extends the LoRA framework by decoupling low-rank adaptation into separate direction and scaling components, which allows more flexible control over feature modulation. This approach enhances expressiveness while maintaining parameter efficiency, leading to improved performance under constrained computational budgets.
    
    \item \textit{Model 10 (BitFit)}~\cite{zaken2021bitfit}: Restricts training to only the bias terms of each layer, minimizing parameter count to an extreme extent. This model emphasizes the surprising effectiveness of minimal adaptation.

    \item \textit{Model 11 (DiffFit)}~\cite{xie2023difffit}: Employs a diffusion-specific PEFT strategy, leveraging architectural insights tailored to generative diffusion models.
\end{itemize}


\subsection{Evaluating Synthesized Images}
We evaluate the synthesized medical images for visual quality and their practical utility for downstream clinical applications. 

\noindent\textbf{Image Generation Quality} Similar to prior work~\cite{saremi2025rl4med,muller2023multimodal}, we assess image quality using standard metrics covering visual fidelity, 
and distributional similarity. We employ the Fr\'echet Inception Distance (FID)~\cite{heusel2017gans} to evaluate the distributional similarity between synthesized and real image collections. 
The Vendi Score~\cite{friedman2022vendi} measures the diversity of generated samples, 
complementing FID’s realism-oriented evaluation. 

Additionally, image-prompt alignment is evaluated at the class level using a pre-trained chest X-ray multi-head Efficient-Net~\cite{tan2019efficientnet} classifier to assess whether the synthesized images accurately reflect their intended diagnostic labels. For each disease condition, a set of 5000 images conditioned on the corresponding textual prompt are generated and passed through the pretrained classifier. Alignment is quantified by measuring the proportion of generated images that are correctly classified into their respective prompted categories, serving as a proxy for semantic faithfulness. This classifier-based evaluation complements distributional metrics by explicitly testing whether disease-specific visual characteristics are preserved and correctly expressed in generated outputs, thus providing a targeted assessment of clinical relevance and prompt adherence.

\noindent\textbf{Usefulness of the synthesized images} We evaluate the practical utility of these images by training classifiers exclusively on synthetic medical data and testing them on real clinical datasets across multiple disease categories. This provides direct evidence of clinical relevance, measured through standard classification metrics such as accuracy. 

\section{Experiments and Results}
\subsection{Dataset and Implementation Details}
We perform experiments on the publicly available CheXpert dataset~\cite {irvin2019chexpert}. All strategies in Table~\ref{tab:fine-tuning-strategies} are fine-tuned on the training set mentioned in Table~\ref{table:data-distribution}. It is important to note that the held-out test set is intentionally made larger than the training set to more rigorously evaluate the generalization performance of the classifier trained on synthesized images. To ensure fair comparison, all methods were fine-tuned on four 80GB H100 GPUs.

\begin{table}[b]
\centering
\caption{Summary of the train and test splits for the six diseases under observation in CheXpert. Note: Individual images can reflect the presence of several concurrent diseases.}
\begin{tabular}{cccc}
\hline
\textbf{Class} & \textbf{Training} & \textbf{Validation} & \textbf{Test} \\ \hline
Cardiomegaly & 3173 & 1195 & 4515 \\
Lung Opacity & 10269 & 4075 & 14658 \\
Edema & 6447 & 2584 & 9210 \\
No Finding & 1801 & 722 & 2591 \\
Pneumothorax & 2196 & 827 & 3027 \\
Pleural Effusion & 9001 & 3505 & 12972 \\ \hline
\end{tabular}
\label{table:data-distribution}
\end{table}

\subsubsection{Metrics Computation}
To compute FID, we use a DenseNet-121~\cite{iandola2014densenet} feature extractor pretrained on chest radiographs (via TorchXRayVision), applied to resized $224\times224$ grayscale images. For the Vendi Score, 1024-dimensional latent features are extracted from the same DenseNet-121 model. Evaluation is performed on a fixed subset of up to 5,000 real and 5,000 synthetic samples per condition. When fewer than 5,000 real samples are available for a given condition, the number of synthetic samples is matched accordingly. This results in a total of 25,133 real and 25,133 synthetic images used for computing the global FID and Vendi Score, sampled from six target conditions in the held-out test set.

\subsection{Results}
We note that certain fine-tuning configurations, such as those involving only the VAE or only the text encoder, and DiffFit (specifically models \# 3, 5, 6, 7 and 11 from Table~\ref{tab:fine-tuning-strategies}), were excluded from detailed analysis due to consistently poor image quality. These models often produced unrealistic or non-medical outputs, limiting their interpretability and usefulness for downstream evaluation.

\begin{table}[t]
\centering
\caption{Evaluating the quality and consistency of the synthesized images. Ca: Cardiomegaly, Lo: Lung Opacity, Ed: Edema, Nf: No Finding, Pn: Pneumothorax, Pe: Pleural Effusion}
\begin{tabular}{cccclllll}
\toprule
\textbf{Models} & \textbf{FID$\downarrow$} & \textbf{Vendi$\uparrow$} & \multicolumn{6}{c}{\textbf{Class Consistency$\uparrow$}} \\
Trainable Component& \multicolumn{1}{l}{} & \multicolumn{1}{l}{} & \multicolumn{1}{l}{Ca} & Lo & Ed & Nf & Pn & Pe \\ \midrule
U-Net & 3.42 & 5.65 & 24.1 & 10.7 & 25.7 & 91.6 & 15.7 & 21.6 \\
U-Net + Text Encoder & 6.57 & 2.79 & 11.0 & 3.1 & 8.2 & 96.9 & 9.99 & 5.8 \\
U-Net + VAE & 7.46 & 2.59 & 12.1 & 1.0 & 0.7 & 98.1 & 48.0 & 2.0 \\
LoRA~\cite{hu2022lora} & 5.65 & 2.64 & 5.0 & 0.5 & 0.5 & 98.9 & 1.0 & 1.8 \\
DoRA~\cite{liu2024dora} & 5.72 & 3.18 & 8.3 & 0.8 & 0.7 & 98.8 & 0.9 & 1.3 \\
BitFit~\cite{zaken2021bitfit} & 5.05 & 5.89 & 4.8 & 12.6 & 9.8 & 86.4 & 13.4 & 5.0 \\ \bottomrule
\end{tabular}
\label{tab:model_quality}
\end{table}

\subsubsection{Qualitative Evaluations} 
\begin{figure}
    \centering
    \includegraphics[scale=0.22]{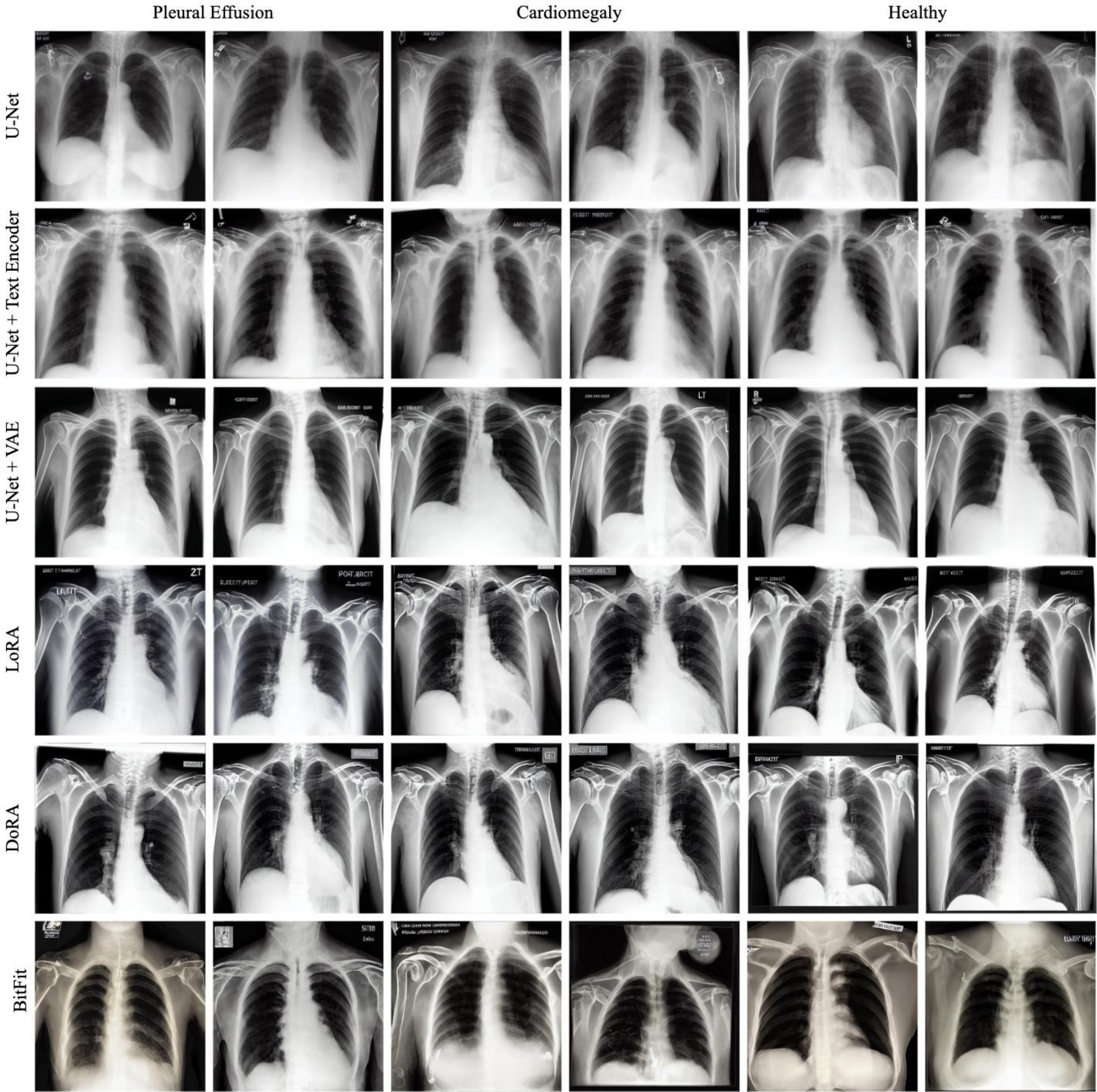}
    \caption{Qualitative comparison of generated chest X-rays across different fine-tuning strategies and three disease categories (Pleural Effusion, Cardiomegaly and Healthy). Each row shows a different fine-tuning method (two samples per disease category). 
    The comparison highlights differences in anatomical plausibility, disease-specific features, and generative fidelity across strategies. Note: Configurations such as those involving only VAE or only text encoder and DiffFit fine-tuning are excluded due to poor image quality.}
    \label{fig:qual_results}
\end{figure}

We present a qualitative comparison of generated chest X-ray images across various fine-tuning strategies in Figure~\ref{fig:qual_results}. The results highlight the effect full fine-tuning, particularly of the U-Net component in the Stable Diffusion architecture, which serves as the core generative module responsible for denoising and image synthesis. Without updating the U-Net, the model struggles to internalize and reproduce medical features accurately. For this reason, models that omit U-Net fine-tuning, such as those only updating the VAE or Text Encoder, are excluded from Figure~\ref{fig:qual_results}, as they tend to produce unrealistic outputs that closely resemble those of the original unadapted Stable Diffusion model. Several parameter-efficient fine-tuning (PEFT) methods yield visually competitive results. Both LoRA and DoRA can reproduce key pathological markers with high visual fidelity, despite training only a small subset of the model parameters. This demonstrates their effectiveness in adapting large generative models to the medical domain with reduced computational cost. In contrast, BitFit, while extremely lightweight, often produces blurrier and less structurally coherent outputs, particularly in areas where fine-grained anatomical detail is critical. Moreover, in some cases, BitFit produces RGB-like images instead of grayscale radiographs, indicating poor adaptation and suggesting that this method may be insufficient for domain-specific fine-tuning in high-stakes settings such as medical imaging. These qualitative findings suggest that while full-component fine-tuning yields the highest visual fidelity, PEFT strategies provide a compelling trade-off between efficiency and image quality, especially for scalable or resource-constrained deployments. 

\begin{table}
\centering
\caption{Accuracy of different model configurations across disease categories. Note: All models are trained and validated on synthetic data and tested on real images.}
\begin{tabular}{l@{\hskip 10pt}c@{\hskip 10pt}c@{\hskip 10pt}c@{\hskip 10pt}c@{\hskip 10pt}c@{\hskip 10pt}c}
\toprule
\textbf{Model} & \textbf{Ca} & \textbf{Lo} & \textbf{Ed} & \textbf{Nf} & \textbf{Pn} & \textbf{Pe} \\
\midrule
U-Net         & 37.4 & 48.8 & 67.8 & 90.9 & 89.4 & 51.5 \\
U-Net + Text Encoder    & 77.9 & 54.0 & 67.8 & 81.0 & 89.4 & 55.3 \\
U-Net + Vae     & 15.7 & 48.8 & 67.5 & 90.9 & 89.4 & 54.7 \\
LoRA~\cite{hu2022lora}         & 35.3 & 47.6 & 60.2 & 75.8 & 86.4 & 53.6 \\
DoRA~\cite{liu2024dora}         & 80.6 & 47.4 & 52.9 & 75.8 & 89.3 & 52.7 \\
BitFit~\cite{zaken2021bitfit}       & 77.3 & 50.9 & 67.8 & 90.9 & 76.4 & 54.5 \\
\bottomrule
\end{tabular}

\label{tab:model_accuracy}
\end{table}
\noindent\textbf{Quantitative Evaluations} Table~\ref{tab:model_quality}, shows trade-offs between fidelity, diversity, and downstream utility across different fine-tuning strategies. Full fine-tuning of the U-Net achieves the best FID of 3.42 and strong class consistency scores, particularly for \textit{No Finding} (91.6) and \textit{Lung Opacity} (25.7), confirming the importance of updating the core generative module. Among PEFT methods, LoRA and DoRA achieve competitive FID scores (5.65 and 5.18, respectively) and maintain high class consistency in major categories like \textit{No Finding} and \textit{Pneumothorax}. 

When evaluated on real test images after training on synthetic data (Table~\ref{tab:model_accuracy}), models fine-tuned on both U-Net and Text Encoder achieve the highest accuracy across nearly all disease categories (e.g., 77.9\% for \textit{Cardiomegaly}, 94.8\% for \textit{No Finding}), highlighting strong generalization. BitFit also performs surprisingly well in this evaluation, suggesting that while it may underperform in generation quality, its generated images still retain enough semantic structure to be informative for downstream classification. LoRA and DoRA also demonstrate robust cross-domain transfer, particularly in high-signal classes such as \textit{Pneumothorax} and \textit{No Finding}, underscoring their utility as efficient yet effective fine-tuning alternatives.

The PEFT methods show computational advantages, with LoRA requiring only 1.59 million trainable parameters compared to the full U-Net training approaches that utilize between 83.7 million and 98.3 million parameters.  Despite this, training times remain comparable, with PEFT methods taking 61-70 seconds per epoch versus 77-177 seconds for full training.

\section{Conclusion}
In this work, we systematically evaluated fine-tuning strategies for adapting Stable Diffusion, a text-to-image foundation model, to high-resolution medical imaging tasks. We compared parameter-efficient methods (LoRA, DoRA, BitFit, and DiffFit) against full U-Net training using a comprehensive evaluation framework that assessed both image quality metrics and downstream task performance. Our results demonstrate that full U-Net training outperforms all parameter-efficient methods across evaluation metrics, establishing it as the optimal approach for high-resolution medical image synthesis when computational resources permit. While parameter-efficient methods successfully address data scarcity challenges in medical imaging domains, the substantial performance gap observed indicates that practitioners should prioritize full U-Net training to achieve maximum diagnostic accuracy. The evaluation methodology developed provides a robust framework for future research by ensuring technical improvements translate to clinical utility. These findings offer clear guidance for deploying foundation models in medical imaging environments and establish performance benchmarks for future parameter-efficient approaches. The work enables informed decisions regarding computational trade-offs in resource-constrained settings while demonstrating the continued superiority of comprehensive network optimization for critical medical applications.

\section{Acknowledgement}
Funding was provided in part by the Natural Sciences and Engineering Research Council of Canada, the Canadian Institute for Advanced Research (CIFAR) Artificial Intelligence Chairs program, Mila - Quebec AI Institute, Google Research, Calcul Quebec, and the Digital Research Alliance of Canada.

\bibliographystyle{plainnat}   
\bibliography{Paper-0025}      

\end{document}